\documentclass{aastex63}

\accepted{ApJ}

\def\um{$\mu$m }

\def\h2o{H$_2$O}

\def\ch4{CH$_4$}

\def\arcs{\ifmmode {''}\else $''$\fi}
\newcommand{\teff}{\mbox{$T_{eff}$}}

\begin{document}

\title{A Survey of 3 to 5.4 $\mu$m Emission From Planetary Nebulae using SOFIA/FLITECAM}

\correspondingauthor{Erin C. Smith}
\email{erin.c.smith@nasa.gov}

\author[0000-0003-0819-4359]{Erin C. Smith} 
\affiliation{NASA Goddard Space Flight Center, Greenbelt, MD 20771}

\author[0000-0002-9632-9382]{Sarah E. Logsdon}
\affiliation{NSF's National Optical-Infrared Astronomy Research Laboratory, Tucson, AZ 85719}

\author{Ian S. McLean}
\affiliation{Department of Physics and Astronomy, UCLA, Los Angeles, CA 90095-1547}

\author{Elizabeth Fletcher}
\affiliation{Department of Physics, Astronomy and Geosciences, Towson University, Towson, MD 21252}

\author[0000-0002-9123-0068]{William D. Vacca}
\affiliation{SOFIA-USRA, NASA Ames Research Center, Moffet Field, CA 94035-1000}

\author{E. E. Becklin}
\affiliation{SOFIA-USRA, NASA Ames Research Center, Moffet Field, CA 94035-1000}
\affiliation{Department of Physics and Astronomy, UCLA, Los Angeles, CA 90095-1547}

\author[0000-0003-0281-7383]{Sachindev Shenoy}
\affiliation{Space Science Institute, Boulder CO}

\author{Maureen Savage}
\affiliation{UCO Lick Observatories, 1156 High St, Santa Cruz, CA 95064}

\author[0000-0001-6350-2209 ]{Ryan T. Hamilton}
\affiliation{Lowell Observatory, Flagstaff, AZ 86001}

\begin{abstract}

Here we present the results of an airborne 3-5.4 \um spectroscopic study of three young, Carbon-rich planetary nebulae IC 5117, PNG 093.9-00.1, and  BD $+$30 3639. These observations were made using the grism spectroscopy mode of the FLITECAM instrument during airborne science operations onboard NASA's Stratospheric Observatory for Infrared Astronomy (SOFIA). The goal of this study is to characterize the 3.3 \um and 5.25 \um PAH dust emission in planetary nebulae and study the evolution of PAH features within evolved stars before their incorporation into new stellar systems in star-forming regions. Targets were selected from IRAS, KAO and ISO source lists, and were previously observed with FLITECAM on the 3-meter Shane telescope at Lick Observatory to allow direct comparison between the ground and airborne observations. We measure PAH emission equivalent width and central wavelength, classify the shape of the PAH emission, and determine the PAH/Aliphatic ratio for each target. The 3.3 \um PAH emission feature is observed in all three objects. PNG 093.9-00.1 exhibits NGC 7027-like aliphatic emission in the 3.4--3.6 \um region while  IC 5117 and BD +30 3639 exhibit less aliphatic structure. All three PNs additionally exhibit PAH emission at 5.25 $\mu$m.

\end{abstract}

\keywords{infrared: general --- techniques: spectroscopic
--- planetary nebulae: general --- stars: AGB and post-AGB --- astrochemistry: PAHs--Observatory: SOFIA}

\section{Introduction} \label{sec:intro}

Due in part to their ubiquity, the broad infrared spectral features at 3.3, 6.2, 7.7, 8.6, 11.3 and 12.7 $\mu$m are an important focus of near and mid-infrared astronomy. Originally referred to as the Unidentified Infrared Bands (UIB), through comparisons with laboratory data, they are generally attributed to the vibrational, bending and stretching modes of the C--H and C--C bonds in Polycyclic Aromatic Hydrocarbons (PAHs) \citep{1975ApJ...200..609G,  1984ApJ...277..623S, 1984AandA...137L...5L}. PAHs are assemblies of benzene-ring like structures, and include molecules such as Pyrene, Naphthalene, and Coronene, which when excited by absorption of a UV photon, relax through emission in the infrared  \citep{2004ApJ...611..928V}. These broad emission features have been observed in a wide range of astronomical objects, including planetary nebulae (PNs), HII regions, proto-stellar clouds and star-forming galaxies \citep[e.g.][] {2004ApJ...604..252P,2004ApJS..154..296H}. This has led to several studies aimed at correlating PAH emission with physical properties of the observed region \citep[e.g.][] {1991ApJ...380..452T, 1996MNRAS.280..924R, 2005MNRAS.362.1199C, 2004ApJ...611..928V}. While the laboratory spectrum of individual PAHs is known, mixing makes identification of individual PAHs in astronomical environments extremely difficult \citep[See section II][]{1989ApJS...71..733A}, resulting in descriptions of overall PAH population characteristics and environmental variations rather than identification of specific PAHs \citep{2002AandA...390.1089P} .  

PAH formation is believed to occur in the last stages of evolution of carbon-rich low- and intermediate-mass stars. \citet{2008ApJ...676..408S} and \citet{1996MNRAS.280..924R} used the 3.3 \um band to determine the minimum C$/$O abundance ratios for PAH emission. They found `cutoff' C$/$O abundance ratios of 0.65 $\pm$ 0.28 and 0.77 $\pm$ 0.5, respectively. \citet{2005MNRAS.362.1199C} similarly used the 7 and 8 $\mu$m bands to find a C$/$O threshold ratio of 0.54 $\pm$ 0.3. This supports the hypothesis that Carbon-rich planetary nebulae produce hydrocarbons including PAHs \citep[See section IV][]{1989ApJS...71..733A}.

The most prominent hydrocarbon band in the 3-5.5\um region is a broad feature centered at 3.3$\mu$m, which sits on top of an extended plateau of emission, extending from $\sim$3.2$\mu$m to $\sim$3.6$\mu$m \citep[][and references therein]{2018A&A...610A..65M} In most PAH emitting objects this is accompanied by a weaker band of emission beginning at 3.4\um and extending to nearly 3.6 $\mu$m. This feature is believed to arise from emission from the C--H stretching of aliphatics, chemical subgroups which do not show the alternating double- and single-bond structure between Carbon atoms that aromatic compounds exhibit \citep{1981MNRAS.196..269D}. Proposed carriers include aliphatic side-groups (such as methyl) attached to the outer carbon molecules of PAH assemblies \citep{1981MNRAS.196..269D,2017NewAR..77....1Y}, and HnPAHs, hydrogenated PAH molecules with additional hydrogen molecules attached to the carbon ring \citep{1996ApJ...472L.127B}. Comparison of 3 $\mu$m spectra to laboratory spectra of methyl-coronene \citep{1996ApJ...458..610J}and HnPAHs \citep{1996ApJ...472L.127B} support both the side-group and HnPAH models. \citet{1997ApJ...474..735S} suggested a mixed chemistry involving both side-groups and HnPAHs could be responsible for the observed 3.4 $\mu$m emission in the Orion Bar. Detailed experimental work on PAH emission in the 3\um region has been published recently by \citet{2018A&A...610A..65M}. These authors show that hydrogenated PAHs form an important fraction of the IR emission carriers in regions with an anomalously strong 3.2 to 3.6\um plateau. For simplicity, in this paper we refer to the 3.4 $\mu$m complex that sits atop the 3.2$\mu$m to 3.6$\mu$m plateau as the `aliphatic feature' and the 3.3 $\mu$m complex that sits above the plateau as the `PAH feature'. 

In addition to the features in the 3.2 to 3.6 \um region, several studies have also noted the presence of PAH bands at 5.25 and 5.7 $\mu$m including on KAO \citep{1989ApJ...345L..59A}, with UKIRT \citep{1996MNRAS.281L..25R}, and with ISO \citep{2009ApJ...690.1208B}. These features have been observed in a small number of planetary nebulae \citep[e.g. NGC 7027, PNG 093.9-00.1, and  BD $+$30 3639][]{1996MNRAS.281L..25R, 2009ApJ...690.1208B}, as well as objects like the Orion Bar \citep{2009ApJ...690.1208B}. \citet{2009ApJ...690.1208B} used observational, laboratory and theoretical data to investigate this feature. They attributed the 5.25 and 5.7 \um emission to combinations of CH stretching and CH in-plane and out-of-plane bending modes and found that the 5.25 and 5.7 $\mu$m features preferentially arise from large, compact PAHs, rather than irregularly shaped molecules.

This study uses guaranteed time observations (GTO) made using the Stratospheric Observatory for Infrared Astronomy (SOFIA) and the FLITECAM (First Light Test Experiment Camera) instrument to investigate the 3-- 5.41 $\mu$m spectra of three young, carbon-rich planetary nebulae.  It has three purposes: investigate the properties of 3.3 $\mu$m PAH emission in these planetary nebulae by expanding the analysis performed by \citet{1996MNRAS.280..924R} and \citet{1989ApJ...341..246C}; examine the 3.4 \um aliphatic emission in these objects and characterize any 5.25 \um PAH emission if detected.

Target selection, the FLITECAM grism mode, and the observations are all described in \S2, data reduction methods are explained in \S3, and results and discussion are presented in \S4. Finally, conclusions are given in \S5.

\section{Observations}

\subsection{Target Selection}

The three objects in this study (IC 5117, PNG 093.9-00.1, and  BD $+$30 3639) are young, carbon-rich planetary nebulae with known 3.3 \um emission \citep{1996MNRAS.280..924R, 2011AJ....141..134S}. Young PNs have only recently begun to exhibit the ionized shells of dust and gas lost during a star's post-AGB phase. The central stars of  young PNs tend to have lower effective temperatures than more evolved PNs \citep{2011AJ....141..134S}. Additionally, young PNs tend to be compact, with significant dust opacities \citep{1994PASP..106..344K}. The study of young PNs allows astronomers to study post-AGB mass loss, to understand the dynamical forces which shape nebular structure, and to investigate the initial stages of dust formation, all of which is made more difficult in older PNs, which can be affected by interactions with the ISM, the influence of the central star wind, and the effects of photoionization \citep{2011AJ....141..134S}. 

 BD $+$30 3639 is a bright, young Planetary Nebulae with a spectral classification of [WC9] \citep{2018A&A...620A..98H, 2011A&A...526A...6W}, and a central star temperature of $\sim$47,000 K \citep{1996A&A...312..167L}. BD $+$30 3639 exhibits 3.3 \um and aliphatic emission as well as 5.25 \um emission \citep{1989ApJ...345L..59A}. IC 5117 is a very compact PN which exhibits bi-polar outflows,  molecular hydrogen emission and aliphatic emission in the 3.4 \um region \citep{1985AJ.....90...49K, 2005MNRAS.362.1199C}. \citet{2001ApJ...563..889H} found an evolutionary age of $\sim$7000 yr for IC 5117. The central star of IC 5117 has been estimated to be $\sim$120,000 K \citep{2001ApJ...563..889H}, substantially hotter than that in BD +30 3639. The high-temperature central star may explain why IC 5117 exhibits high-energy forbidden line emission \citep{2001ApJ...563..889H}. PNG 093.9-00.1 (also called IRAS 21282$+$5050) is a young, low excitation PN with a [WC11] spectral type \citep{1993ApJ...411..266M, 1993ApJ...408..586K} and a $\sim$28,000 K central star \citep{1998A&A...330..265L}. It exhibits an abnormally large infrared excess and multipolar morphology \citep{2019Ap&SS.364...32H}. PNG 093.9-00.1 is believed to be evolutionarily younger than both IC 5117 and BD $+$30 3639, with an evolutionary age of $\sim$5000 yr \citep{1989ApJ...345L..55S}. 
 
IC 5117, PNG 093.9-00.1 and BD $+$30 3639  have been previously observed by FLITECAM in a ground-based survey using the Shane 3-meter telescope at Lick Observatory, Mt. Hamilton, CA \citep{2008ApJ...676..408S}.  The \citet{2008ApJ...676..408S} survey measured the 3.3 \um PAH emission in all three targets, but atmospheric water vapor made examination of the 3.4 \um aliphatic feature difficult. Likewise, the 5.25 \um feature was unobservable from the ground. These three objects were selected for airborne observations because the strength of their PAH emission made them optimal targets for detection of aliphatic and 5.25 \um emission if present.

\subsection{Instrumentation}

Observations of all three objects were carried out with FLITECAM. FLITECAM is a 1--5.41 \um camera developed at the UCLA Infrared Laboratory under the leadership of Dr. Ian McLean for NASA's SOFIA project \citep{2006SPIE.6269E.168M}. SOFIA is a modified Boeing 747-SP airplane with a 2.5-meter f/19.6 bent-Cassegrain telescope operating at altitudes up to 45,000 ft and therefore above 99\% of the atmosphere's water vapor content \citep{2002SPIE.4486...77W}. Performance and technical details of the instrument are described in \citet{2006SPIE.6269E.168M}. FLITECAM was commissioned aboard SOFIA in 2015 \citep{2014SPIE.9147E..06S,2014SPIE.9147E..2YL, 2016SPIE.9908E..0BL}, but because FLITECAM operates in the near-infrared, and because the f/17 optics of the Lick 3-meter telescope provides a plate scale similar to that of SOFIA, FLITECAM was also commissioned in October 2002 at Lick Observatory in a ground based configuration \citep{2003ApJ...597..555M}. FLITECAM was retired from SOFIA operations in 2018.

FLITECAM utilizes large refractive optics to inscribe an $\sim$8 arc minute diameter FOV on a 1024x1024 InSb ALADDIN III detector with a plate scale of 0.475 arc seconds per pixel. FLITECAM was designed as an imaging camera, but also has a suite of three grisms which allow for nearly full coverage of the 1-5.41\um range at a resolution of R $\sim$1300-1800. The instrument can be operated in either the up-looking position (Lick) or the horizontal position (SOFIA). Although these measurements were made in the FLITECAM-only mode, FLITECAM can be co-mounted with another SOFIA instrument, HIPO (High-speed Imaging Photometer for Occultations) to provide simultaneous coverage of the optical and near-infrared bands \citep{2014SPIE.9147E..0HD, 2004SPIE.5492..592D, 2014SPIE.9147E..2YL}. 

FLITECAM features a dual filter wheel which houses standard broadband filters, several narrow band filters plus a suite of three direct-ruled KRS-5 grisms and their appropriate order-sorting filters. Each grism can be used in one of three orders giving a total of 9 settings to nearly cover the entire 1--5.41 $\mu$m region. The KRS-5 grisms provide a moderate resolution spectroscopy mode in conjunction with fixed slits of either 1\arcs~ or 2\arcs~ width and 60\arcs~ total length to yield resolving powers of R$\sim$1800 and 1300 respectively \citep{2006SPIE.6269E..50S}. See Table 1 for details of the FLITECAM grism mode. The 3-3.365\um range was observed with the LM order sorting filter and the C grism. This combination is referred to as the LMC band. The 3.365-4.1\um range was observed with the  LM order sorting filter and the B grism, referred to as the LMB band. There is an overlap between the LMC and LMB bands starting near the 3.3\um feature peak, and continuing to 3.4\um. The transition between LMC and LMB data was chosen to fall at  3.365\um to avoid both the 3.3 and 3.4 micron features. The 4.4-5.41\um range was observed with the  LM order sorting filter and the A grism. This is referred to as the LMA band.

\subsection{Observations}

Observations were obtained over the course of three separate observing flights onboard SOFIA in October 2016. Spectroscopic observations were made in a standard ABBA nod pattern. Individual exposures ranged from 30-80s per observation, depending on target and wavelength band.The object was nodded between two positions, designated A and B, separated by $\sim$20$''$ on the $\sim$60$''$ long entrance slit of FLITECAM. Signal-to-noise ratios were typically greater than 10 (10\%) per resolution element. Seeing conditions were $\sim2\farcs0-3\farcs0$ and a slit width of 2$\farcs$0 ($\sim$4 pixels) was used for all observations. Each object was observed in at least three grism modes, providing coverage from $\sim$2.8 $\mu$m to 5.41 $\mu$m, except for the 4.2-4.4 $\mu$m region, which is opaque even at SOFIA altitudes. PNG 093.9-00.1 was additionally observed in the 1.8-2.4$\mu$m band. Standard infrared stellar calibrators were observed for wavelength and flux calibration. A summary of all observations can be found in Table 2.

\section {Data Reduction}
Data from all three objects was reduced using REDUX, SOFIA's facility pipeline version 1.2.0 \citep{2015ASPC..495..355C}. REDUX used the FSpextool software package, a version of the NASA Infrared Telescope Facility's SpeX instrument data reduction software, Spextool \citep{2004PASP..116..362C}, that was specifically optimized for FLITECAM to extract, flat-field, and linearize the FLITECAM spectra \citep{2015ASPC..495..355C, 2004PASP..116..362C, 2004PASP..116..352V}. For more detail see the FLITECAM Guest Observer (GO) Data Handbook. 
\footnote{https://www.sofia.usra.edu/sites/default/files/USpot\_DCS\_DPS/Documents/FLITECAM\_GO\_Handbook.pdf}

Because SOFIA data were still subject to telluric contamination, our near-infrared (NIR) spectra needed to be corrected for atmospheric, as well as instrumental effects. FSpextool achieved this by applying the standard ground-based near-infrared (NIR) astronomy correction approach, that is by observations of a telluric standard star, generally an A0V star \citep{2003PASP..115..389V}. The corrected standard star spectra was reduced in the same manner as the science target data, then compared to a scaled Vega model to create a correction curve. In ground-based astronomy telluric standards and science targets usually have been observed at similar airmasses, but due to the unique operational aspects of airborne observations, this was not always possible on SOFIA. The effects of this airmass mismatch were removed by computing the appropriate telluric absorption model spectra for both the telluric standard and the target using the ATRAN code \citep{1992nstc.rept.....L}, then re-binning the computed model spectra to match FLITECAM's spectral resolution. The final telluric correction curve was calculated from the ratios of the telluric calibrator's computed telluric model spectrum to the science target's computed telluric model spectrum. This final correction curve was then applied to the science spectrum to produce a flux-calibrated, telluric corrected spectrum of the target \citep{2015ApJ...804...66V}. 

\section{Results and Discussion}

\subsection{Overview}

The 3.0 to 4.0 $\mu$m and 4.5 to 5.41 $\mu$m spectra for each object are presented in Figures 1 (BD $+$30 3639), 2 (IC 5117) and 3 (PNG 093.9-00.1). The difficulty of positioning the slit exactly in the same place on an extended object accounts for the slight mismatch in continuum levels from one grism observation to the next. The S/N for the 4.5--5.41 \um region was lower than the 3 \um regime due to atmospheric and thermal backgrounds, so data were smoothed using a 5 pixel wide Gaussian filter which reduced the spectral resolution to $\sim$700 in the 4.2-5.4 region. Atomic line emission from Hydrogen was apparent in IC 5117 and BD $+$30 3639, with a tentative detection of Brackett-alpha emission in PNG 093.9-00.1. Additionally, IC 5117 showed [Mg IV] emission. See Table 3.

As expected, all three objects exhibit the 3.2-3.6\um plateau, with the 3.3\um PAH complex and the 3.4 \um aliphatic complex emission evident above the 3.2 to 3.6\um plateau. In PNG 093.9-00.1 and BD $+$30 3639 we also clearly detected the PAH feature at 5.25 $\mu$m. Although weaker, this feature was also detected in IC 5117. The 5.7 $\mu$m feature associated with 5.25\um emission is beyond FLITECAM's wavelength range. Before analyzing the 3.3 micron PAH feature, we smoothed the spectra with a five-pixel wide Gaussian filter and fit the Pfund-$\delta$ Hydrogen emission line at 3.296 $\mu$m, which overlaps with the peak of the expected 3.3 \um PAH feature, with a Gaussian. The Pfund-$\delta$ feature was removed using IDL's Gaussfit routines and MPFIT \citep{2009ASPC..411..251M}\footnote{http://purl.com/net/mpfit}. The continuum for the 3.3 and 3.4 \um features was fit using polyfit, with continuum points at 3.0 \um $\pm$ 0.025 \um and 3.8 \um $\pm$ 0.025$\mu$m. We then calculated the equivalent width and flux of each feature. Using MPFIT and a modified version of XGAUSSFIT \citep{2001fitr.book.....L} we fit the 3.4\um aliphatic feature with four overlapping Gaussians at 3.4, 3.46, 3.51 and 3.56 $\mu$m, as identified by \citet{1990AandAS...83..337J}. The 3.3 \um feature was fit with a Gaussian simultaneously with the 3.4 \um feature. The 3.3 and 3.4 \um results are presented in Table 4 and discussed below. The 5.25 \um feature was also fit with a Gaussian using MPFIT, with results in Table 5. 

\subsection{3.3 $\mu$m Aromatic Feature}

\citet{2004ApJ...604..252P} used the full width at half maximum (FWHM), peak wavelength and overall shape of the 6.2, 7.7 and 8.6 $\mu$m PAH spectra to classify the emission into three groups. This approach (though not the definitions of the classification groups) was extended to the to the 3.3 $\mu$m and 11.2 $\mu$m bands by several groups \citep[e.g.][]{2004ApJ...611..928V}.  The 3.3 \um classification scheme uses band symmetry, FWHM and central wavelength to categorize emission into three main classes, identified as Class A$_{\rm 3.3}$, B1$_{\rm 3.3}$ and B2$_{\rm 3.3}$ \citep{2004ApJ...611..928V}. Class A$_{\rm 3.3}$ exhibits symmetric emission centered at 3.290 $\mu$m with a FWHM of 0.040 $\mu$m, and is exhibited by most objects, including planetary nebulae. Class B1$_{\rm 3.3}$ and Class B2$_{\rm 3.3}$ exhibit asymmetric emission with a FWHM of 0.037$\mu$m. Class B1$_{\rm 3.3}$ has a peak wavelength of 3.293$\mu$m, while Class B2$_{\rm 3.3}$ peaks at 3.297$\mu$m \citep{2004ApJ...611..928V}. \citet{1991ApJ...380..452T} used similar criteria to categorize 3.3 $\mu$m emission, with their class 1 equivalent to van Diedenhoven's Class A$_{\rm 3.3}$. 

Previous studies have found little variation in the classification of 3.3 \um emission between different object classes, even though longer wavelength PAH emission can show much greater variation \citep{2008ApJ...676..408S, 2004ApJ...611..928V, 2005MNRAS.362.1199C,2002AandA...390.1089P, 1991ApJ...380..452T}. This may be because while the 6.2 and 7.7 $\mu$m features arise from stretching in the C--C bonds in the PAH molecules, the 3.3 and 11.2 $\mu$m features are due to stretching and out-of-plane bending modes of the C--H bonds  \citep[See Table 1,][]{1989ApJS...71..733A}, which may lead to smaller variation in emission in the 3.3 and 11.2 $\mu$m features.  

The central wavelength and FWHM of the 3.3 $\mu$m feature were calculated by modeling the continuum at 3.0 \um and 3.8 \um, and then removing the contribution of the Pfund $\delta$ Hydrogen emission line at 3.296 \um (see section 4.1). Following the original classification papers \citep{2002AandA...390.1089P, 2004ApJ...611..928V, 1991ApJ...380..452T} we fit a symmetric Gaussian to the 3.3\um emission feature using MPFIT \citep{2009ASPC..411..251M}. It should be noted that the 3.3\um emission is not a Gaussian, and was only approximated with a Gaussian for consistency with past classification schemes \citep[e.g.][]{2002AandA...390.1089P}. After calculating the central wavelength and FWHM, we applied the classifications used by \citet{2004ApJ...611..928V} to the 3.3 \um emission of all three PN.  See Table 4. As expected, the three planetary nebulae exhibit class A$_{\rm 3.3}$ behavior, as is consistent with previous surveys \citep{2004ApJ...611..928V,1996MNRAS.280..924R, 1991ApJ...380..452T}. 

SOFIA does not have a field de-rotator, and FLITECAM has a single fixed slit for grism observations,  so this survey did not sample multiple positions across the nebulae. Therefore, this technique would not detect subtle, spatially-dependent variations in 3.3 $\mu$m emission within the nebulae. FLITECAM's two arcsecond slit used for these observations does not sample the entirety of PNG 093.9-00.1 and BD $+$30 3639 or IC 5117. When slit losses are taken into account, the ground-based 3.3 \um fluxes reported in \citet {1989ApJ...341..246C} are $\sim$70$\%$ of our 3.3 \um fluxes for BD $+$30 3639 and IC 5117. (See Table 4).

\subsection{Aliphatic Features}
The 3.4\um aliphatic emission complex has been observed as a band of emission sitting on top of the 3.2-3.6\um plateau in most objects exhibiting 3.3 $\mu$m PAH emission, including HII regions \citep[e.g.][]{1997ApJ...474..735S}, proto-planetary nebulae \citep[e.g.][]{2007ApJ...662.1059H}, external galaxies \citep[e.g.][]{2000ApJ...545..701I} and planetary nebulae \citep[e.g.][]{1985ApJ...292..500G}. The aliphatic emission feature is in general weaker than the PAH feature except in carbon-rich proto-planetary nebulae \citep{1990A&A...235L...9G}, where the 3.4 $\mu$m plateau can dominate the 3 $\mu$m spectrum. \citet{2001ApJ...554L..87K} have proposed that the apparent evolution from aliphatic-dominated to aromatic-dominated emission results from the destruction of aliphatics by the increasing UV field as objects evolve from UV weak proto-planetary nebulae to UV-hard planetary nebulae. The aliphatic band is sometimes observed as a series of individual peaks, rather than as a smooth feature above the 3.2-3.6\um plateau. \citet{1990AandAS...83..337J} identified these bands as occurring at 3.4, 3.46, 3.51 and 3.56 $\mu$m.

All three objects in this data set do exhibit emission above the 3.2-3.6\um plateau in the 3.4 to 3.6 \um region (see Figures 1, 2 and 3). PNG 093.9-00.1 exhibited the most intense aliphatic emission in comparison with its PAH emission. Additionally, in the 3.4 to 3.6 \um region, PNG 093.9-00.1 shows evidence of the series of peaks occurring at 3.4, 3.46, 3.51 and 3.56 $\mu$m identified by \citet{1990AandAS...83..337J}. Aliphatic emission in IC 5117 and BD +30 3639 is much weaker in comparison to their PAH emission, and does not appear to exhibit the structure visible in PNG 093.9-00.1, instead appearing as a sharp rise at 3.4 $\mu$m which trails off to continuum levels at $\sim$ 3.6 $\mu$m. The lack of structure may be due to low S/N, rather than the emission itself.  The equivalent width of this overall feature was calculated (see Table 4) and compared to the equivalent width of the aromatic 3.3 \um emission for each object. PNG 093.9-00.1 shows a higher aliphatic-to aromatic ratio than the other objects. Such a result is not unexpected as PNG 093.9-00.1 is a young object, exhibits multipolar morphology (with a binary central object), and is known to have stronger than expected 3.4 \um emission \citep{2019Ap&SS.364...32H,1993ApJ...408..586K,1989ApJ...341..246C}. The PN's high C/O ratio may also have influence on the strength of its aliphatic feature \citep{1998AandA...329..691K}. 

\subsection{The 5.25 \um feature}
The 5.25 $\mu$m feature was first detected by \citet{1989ApJ...345L..59A} in KAO observations, and has been studied in a small number of objects, including HII regions and planetary nebulae. Our observations show 5.25 \um emission in all three Planetary Nebulae in our sample. While 5.25 \um emission has been reported in BD +30 3639 and PNG 093.9-00.1\citep{1989ApJ...345L..59A, 1996MNRAS.281L..25R}, it has not been measurably detected in IC 5117 prior to this work. The 5.25 \um feature was measured by smoothing with a 5-pixel Gaussian, then fitting with MPFIT \citep{2009ASPC..411..251M}. The central wavelength, equivalent width and flux of the feature in each object is listed in Table 5.

The 5.25 $\mu$m feature is generally associated with another feature at 5.7 $\mu$m, which is beyond FLITECAM's wavelength range. The 5.25 and 5.7 $\mu$m features preferentially arise from large, compact PAHs, rather than irregularly shaped molecules and is attributed primarily to combinations of CH stretching and CH in-plane and out-of-plane bending modes \citep{2009ApJ...690.1208B}.

\subsection{4.4 and 4.6 \um Deuterated PAH features}
Using ISO Short Wavelength Spectrometer observations of the Orion Bar and M17, \citet{2004ApJ...604..252P} detected broad emission features at 4.4 and 4.6 $\mu$m which were consistent with predictions for the C--D stretching modes in deuterated aromatics and aliphatics \citep{1997JPCA...101.2414B,2004ApJ...614..770H}. Other surveys have attempted to detect emission from deuterated PAHs (PADs) and aliphatics in the 4.4 to 4.6 \um region \citep{2011EAS....46...55O,2014ApJ...780..114O,2016A&A...586A..65D}, but have generally found very weak PAD emission in only a handful of HII regions. This suggests deuteration of PAHs is uncommon \citep{2016A&A...586A..65D}. In our data set part of the 4.4 $\mu$m PAD feature is obscured by telluric absorption (even at stratospheric altitudes), but the 4.65 $\mu$m feature is not clearly detected in any of the three sources (see Figures 1, 2  and 3). This is consistent with other investigations of PAD emissions which found them to be rare in HII regions, and indicates they may be even rarer in planetary nebulae \citep{2016A&A...586A..65D}. 

\subsection{Discussion}
While their appearance is very similar, the 3--5.41 \um emission in our three objects varies considerably. PNG 093.9-00.1  shows almost no emission lines, with only a potential detection in Brackett-$\alpha$. BD +30 3639 and IC 5117 both show several very strong Hydrogen emission lines, and we also detect forbidden line emission from [Mg IV] in IC 5117 at 4.49\um. Infrared forbidden line emission, which has been observed in planetary nebulae at a lower spectral resolution \citep[e.g.][]{2001ApJ...563..889H}, peaks at high electron density $\sim$10$^7$cm$^-$$^3$. The high central star temperature of IC 5117 \citep[$\sim$120,000;][]{2001ApJ...563..889H} may explain not only the forbidden line emission, but could also explain the relative weakness of the PAH, aliphatic and 5.25\um features. In comparison, the nebula in our sample with the lowest temperature central star \citep[\teff$\sim$28,000 K;][]{1998A&A...330..265L}, PNG 093.9-00.1, is dominated by dust emission at 3.3$\mu$m, 3.4$\mu$m, and 5.25$\mu$m. While all three PNs show PAH and aliphatic emission, PAH emission in PNG 093.9-00.1 is slightly different from IC 5117 and BD +30 3639, with a 3.3 \um central emission wavelength slightly blue-shifted from the more evolved PNs. Additionally, the shape of the PAH emission in PNG 093.9-00.1 is more asymmetric. The underlying structure of the aliphatic emission is also more apparent in PNG 093.9-00.1, with peaks at 3.4$\mu$m, 3.46$\mu$m, 3.51$\mu$m, and 3.56$\mu$m. Even the ratio of the PAH to aliphatic emission is different in this object, with the aliphatic emission in PNG 093.9-00.1 stronger in comparison to its 3.3 \um emission than the other objects. These differences may be due to the age, the temperature of the central star, or the individual morphology of the nebula, or, most likely, a combination of characteristics. A study of the spatial variations within these PNs would potentially uncover additional emission variations due to nebular conditions.

\section{Conclusions}

Using the grism spectroscopy mode of FLITECAM aboard SOFIA, we obtained 3--5.41 $\mu$m spectra of three planetary nebulae IC 5117 and BD +30 3639 and PNG 093.9-00.1. Because SOFIA flies above 99\% of Earth's water vapor, this allowed investigation of not just the bright 3.3 $\mu$m PAH emission, but also the fainter 3.4--3.6$\mu$m aliphatic emission feature as well as the faint 5.25 $\mu$m feature. From these data the 3--5.41 $\mu$m spectrum of PNG 093.9-00.1 does show some minor variations from IC 5117 and BD +30 3639, appearing to exhibit relatively more aliphatic chemistry than the other PNs.

(1) All three objects showed PAH emission, with FWHM and central wavelength consistent with that found by \citet{1991ApJ...380..452T}, and corresponding to Class A$_{\rm 3.3}$ emission.

(2) Aliphatic emission was observed in the 3.4--3.6 $\mu$m range of all three targets, with PNG 093.9-00.1 showing strong aliphatic emission in comparison to its aromatic (3.3 $\mu$m) emission. The observed differences may be due to morphology, central star temperature, or age, but also could be due to spatial variations within the objects. An investigation into the spatial variation of PAH and aliphatic emission in planetary nebulae could reveal important clues about PAH formation and evolution.

(3) No emission from deuterated PAHs was detected.

(4) The 5.25$\mu$m feature was detected in all three objects.\\

\acknowledgements 
We thank the anonymous referee for the helpful review of this paper. The authors would also like to thank the FLITECAM team and the SOFIA project for their dedication and expertise in enabling these observations. This research is based on observations made with the NASA/DLR Stratospheric Observatory for Infrared Astronomy (SOFIA). SOFIA is jointly operated by the Universities Space Research Association, Inc. (USRA), under NASA contract NNA17BF53C, and the Deutsches SOFIA Institut (DSI) under DLR contract 50 OK 0901 to the University of Stuttgart. [Financial support for this work was provided by NASA {through an award issued by USRA.}] 
\facility{SOFIA (FLITECAM)}
\software{IDL, ATRAN code \citep{1992nstc.rept.....L}, MPFIT \citep{2009ASPC..411..251M}, REDUX \citep{2015ASPC..495..355C}, Spextool \citep{2004PASP..116..362C}, XGAUSSFIT \citep{2001fitr.book.....L}}

\clearpage

\pagestyle{empty}

% Grism mode Summary (table 1)
\begin{deluxetable}{lcccccccc}

\tablewidth{0pt}\tablecaption{\bf FLITECAM GRISM MODE SUMMARY \label{tbl-1}}
\tablehead{
\colhead{Grism}	&	\colhead{Lines/mm}	&	\colhead{Order}	&	\colhead{OSF}	&	\colhead{Lambda start}	&	\colhead{Lambda central}	&	\colhead{Lambda end}	&	\colhead{Measured R}	\\
\colhead{}	&	\colhead{}	&	\colhead{}	&	\colhead{}	&	\colhead{($\mu$m)}	&	\colhead{($\mu$m)}	&	\colhead{($\mu$m)}	&	\colhead{}	}
\startdata
A	&	162.75	&	1	&	LM	&	4.395	&	4.96	&	5.533	&	1680	\\
A	&	162.75	&	2	&	Klong	&	2.216	&	2.5	&	2.784	&	1690	\\
A	&	162.75	&	3	&	Hwide	&	1.497	&	1.69	&	1.877	&	1710	\\
B	&	217	&	1	&	LM	&	3.307	&	3.73	&	4.16	&	1780	\\
B	&	217	&	2	&	Hwide	&	1.649	&	1.86	&	2.076	&	1750	\\
B	&	217	&	3	&	J	&	1.14	&	1.28	&	1.424	&	1720	\\
C	&	130.2	&	2	&	LM	&	2.756	&	3.11	&	3.467	&	1670	\\
C	&	130.2	&	3	&	Kwide	&	1.872	&	2.11	&	2.346	&	1650	\\
C	&	130.2	&	4	&	H	&	1.445	&	1.62	&	1.801	&	1640	\\
\enddata
\tablecomments{See \citet{2016SPIE.9908E..0BL} and \citet{2006SPIE.6269E..50S} for more detail. Resolution measured for 2 arcsecond slit. LMA resolution determined from in-flight observations, other spectral resolutions determined at Lick Observatory.}

\end{deluxetable}
% Table of Observations (table 2)
\begin{deluxetable}{lccccccc}

\tablewidth{0pt}\tablecaption{\bf TARGET LIST AND OBSERVING LOG \label{tbl-2}}
\tablehead{
\colhead{Object} &
\colhead{Spectral} &
\colhead{\teff} &
\colhead{Size (FWHM)} &
\colhead{Obs} &
\colhead{3-3.4 $\mu$m } &
\colhead{3.4-4.1 $\mu$m } &
\colhead{4.4-5.41 $\mu$m } \\
\colhead{Name} &
\colhead{Type} &
\colhead{(K)} &
\colhead{(arcsec)} &
\colhead{Date} &
\colhead{time (s)} &
\colhead{time (s)} &
\colhead{time (s)}
}
\startdata
BD $+$30 3639     & [WC9] & 47000 &5.15 & 2016 Oct 14	 & 360 	& 360 	& 900 \\
IC 5117	& [WR]	&120000 &2.95	& 20016 Oct 19 & 960 	& 1440 	& 960 \\
PNG093.9$-$00.1    & [WC11] & 28000 &3.96 	& 2016 Oct 20 & 480 	& 720 	& 600 \\
\enddata
\tablecomments{spectral types from Weidmann \& Gamen (2011); \teff from: BD $+$303639: Leuenhagen et al. (1996), IC 5117: Hyung et al.( 2001), PNG093.9$-$00.1: Leuenhagen \& Hamann(1998); Angular Size from FLITECAM K-band images}

\end{deluxetable}

\begin{deluxetable}{lccccc}
\tablewidth{0pt} \tablecaption{\bf ATOMIC LINE EMISSION \label{tbl-3}}
\tablehead{ 
\colhead{Object} &
\colhead{Pfund-$\epsilon$ Flux } &
\colhead{Pfund-$\delta$ Flux } &
\colhead{Pfund-$\gamma$ Flux } &
\colhead{Brackett-$\alpha$ Flux } &
\colhead{Pfund-$\beta$ Flux }\\
\colhead{Name} &
\colhead{(10$^-$$^1$$^5$W/m$^2$)} &
\colhead{(10$^-$$^1$$^5$W/m$^2$)} &
\colhead{(10$^-$$^1$$^5$W/m$^2$)} &
\colhead{(10$^-$$^1$$^5$W/m$^2$)} &
\colhead{(10$^-$$^1$$^5$W/m$^2$)}
}
\startdata
BD $+$30 3639    & 2.50$\pm$0.05       & 5.1$\pm$0.3     & 5.64$\pm$0.04   & 41.53$\pm$0.04   &  6.26$\pm$0.31 \\
IC 5117                 & 0.25$\pm$0.09      & 0.56$\pm$0.2     & 0.75$\pm$0.01   & 4.76$\pm$0.01    & 1.17$\pm$0.21 \\
PNG 093.9$-$00.1  & - & - & -  & 0.32$\pm$0.05 &  -   \\
\enddata
\tablecomments{IC 5117 also showed emission in [Mg IV], with a flux of 4.1$\pm$0.18(10$^-$$^1$$^5$W/m$^2$) Flux has been corrected for slit loss assuming the angular size in Table 2 and a 2 arcsecond FLITECAM slit. Extraction width (arcsec): BD: 8.2 (3-3.35$\mu$m), 8.4 (3.35-4.1$\mu$m), 20.1 (4.2-5.41$\mu$m); IC: 4.3 (3-3.35$\mu$m), 3.7 (3.35-4.1$\mu$m), 10.0 (4.2-5.41$\mu$m);PNG: 6.1 (3-3.35$\mu$m), 5.8 (3.35-4.1$\mu$m), 5.4 (4.2-5.41$\mu$m)}

\end{deluxetable}

\begin{deluxetable}{lccccccc}
\tablewidth{0pt} \tablecaption{\bf PAH AND ALIPHATIC RESULTS \label{tbl-4}}
\tablehead{ 
\colhead{Object} &
\colhead{3.3$\mu$m Peak} &
\colhead{3.3$\mu$m FWHM} &
\colhead{3.3$\mu$m} &
\colhead{3.3$\mu$m EW} &
\colhead{Aliphatic EW} &
\colhead{Ratio}&
\colhead{3.3$\mu$m Flux}\\
\colhead{Name} &
\colhead{($\mu$m)} &
\colhead{($\mu$m)} &
\colhead{Class} &
\colhead{($\mu$m)} &
\colhead{($\mu$m)} &
\colhead{3.4/PAH} &
\colhead{(10$^-$$^1$$^5$W/m$^2$)}
}
\startdata
BD $+$30 3639 & 3.292$\pm$0.002 & 0.042$\pm$0.002 & A & 0.204$\pm$0.005  & 0.049$\pm$0.003 & 0.240 & 88.3$\pm$0.07 \\ 
IC 5117                 & 3.292$\pm$0.004 & 0.041$\pm$0.002         & A         & 0.096$\pm$0.007    & 0.027$\pm$0.005 & 0.281 & 11.9$\pm$0.06 \\  
PNG 093.9$-$00.1& 3.290$\pm$0.002 & 0.044$\pm$0.003      & A         & 0.302$\pm$0.002   & 0.142$\pm$0.002 & 0.470  & 39.3$\pm$0.02 \\
\enddata
\tablecomments{ Flux has been corrected for slit loss assuming the angular size in Table 2 and a 2 arcsecond FLITECAM slit. Extraction width (arcsec): BD: 8.2 (3-3.35$\mu$m), 8.4 (3.35-4.1$\mu$m); IC: 4.3 (3-3.35$\mu$m), 3.7 (3.35-4.1$\mu$m); PNG: 6.1 (3-3.35$\mu$m), 5.8 (3.35-4.1$\mu$m). Cohen et al. reported 3.3 \um fluxes for of 62.8(10$^-$$^1$$^5$W/m$^2$) for BD$+$30 3639 and 8.59(10$^-$$^1$$^5$W/m$^2$) for IC 5117}

\end{deluxetable}

\clearpage

\thispagestyle{empty}
\begin{deluxetable}{lccccccc}
\tablewidth{0pt} \tablecaption{\bf 5.25 MICRON RESULTS \label{tbl-5}}
\tablehead{ 
\colhead{Object} &
\colhead{Extraction} &
\colhead{5.25$\mu$m Peak } &
\colhead{5.25$\mu$m FWHM } &
\colhead{5.25$\mu$m EW} &
\colhead{5.25$\mu$m Flux}  \\
\colhead{Name} &
\colhead{Width(arcsec)} &
\colhead{($\mu$m)} &
\colhead{($\mu$m)} &
\colhead{($\mu$m)} &
\colhead{(10$^-$$^1$$^5$W/m$^2$)}
}
\startdata
BD$+$30 3639   & 20.1& 5.265$\pm$0.01 & 0.12$\pm$0.05 & 0.06$\pm$0.01 & 23.6$\pm$2.7  \\ 
IC 5117                  & 10.0 &5.269$\pm$0.02 & 0.09$\pm$0.05 & 0.05$\pm$0.03 & 3.6$\pm$0.9 \\ 
PNG 093.9$-$00.1     &    5.4   & 5.267$\pm$0.01 & 0.15$\pm$0.03 & 0.16$\pm$0.01 & 23.8$\pm$0.5 \\
\enddata
\tablecomments{Flux has been corrected for slit loss assuming the angular size in Table 2 and a 2 arcsecond FLITECAM slit.}

\end{deluxetable}

\clearpage

\thispagestyle{empty}

%Figure 1: BD +30 3639 spectrum
\begin{figure}[hbt!]
\epsscale{1.0}
\plotone{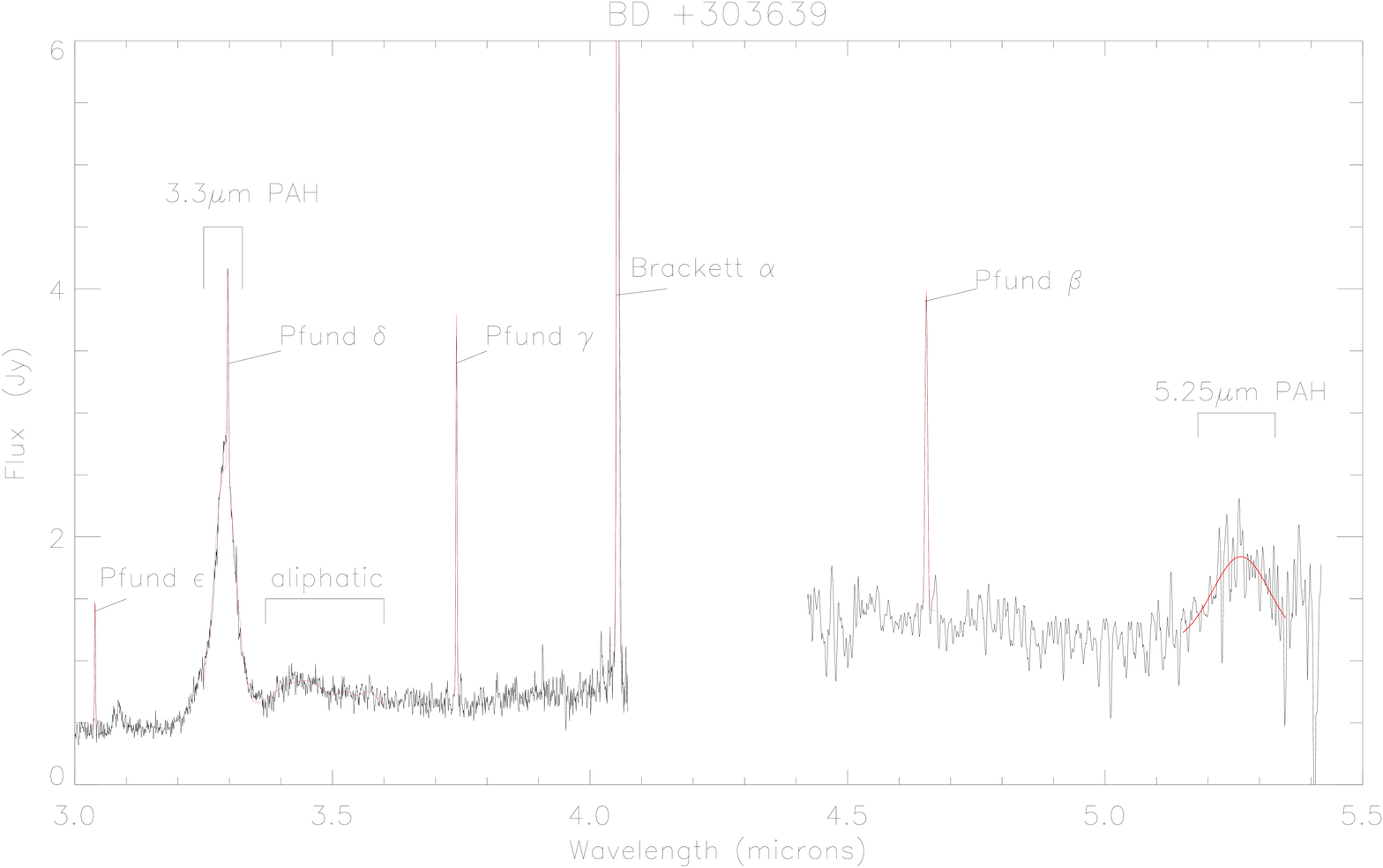}
\caption{Spectrum of BD $+$30 3639 obtained by SOFIA/FLITECAM with a spectral resolution $\sim$1300 from 3 to 4.1$\mu$m and $\sim$700 from 4.2 to 5.4$\mu$m. PAH, Aliphatic and Atomic emission features are labeled. Calculated emission fits are in red.}.\label{fig2}
\end{figure}

%Figure 2: IC 5117 spectrum

\begin{figure}[htb!]
\epsscale{1.0}
\plotone{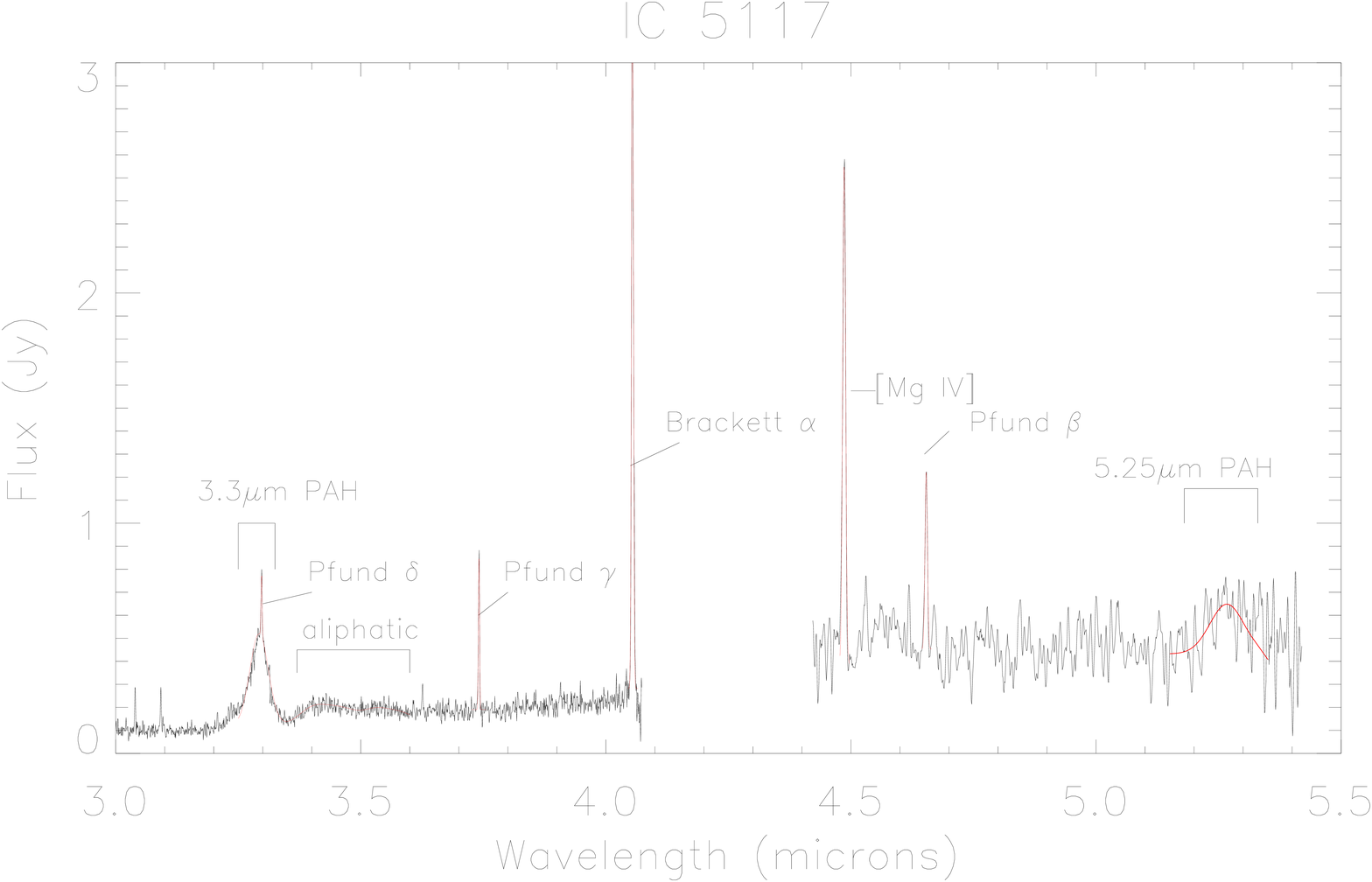}
\caption{Spectrum of IC 5117 obtained by SOFIA/FLITECAM with a spectral resolution $\sim$1300 from 3 to 4.1$\mu$m and $\sim$700 from 4.2 to 5.4$\mu$m. PAH, Aliphatic and Atomic emission features are labeled. Calculated emission fits are in red.}.\label{fig1} 
\end{figure}

%Figure 3: PNG 093.9-00.1 spectrum

\begin{figure}[t!]
\epsscale{1.0}
\plotone{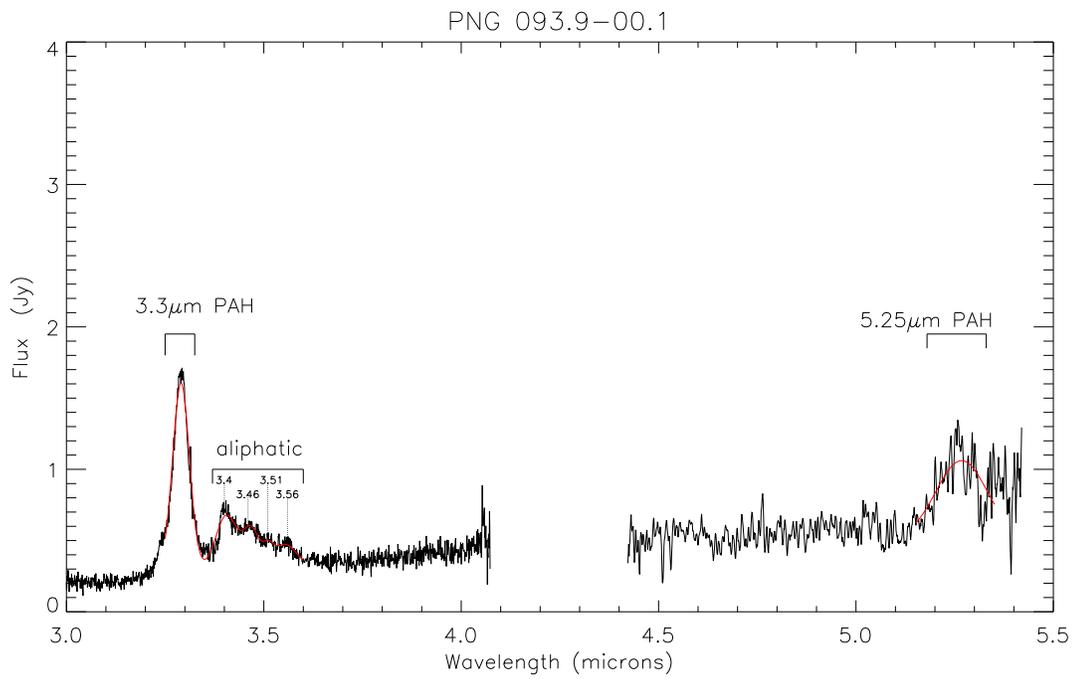}
\caption{Spectrum of PNG 093.9-00.1 obtained by SOFIA/FLITECAM with a spectral resolution $\sim$1300 from 3 to 4.1$\mu$m and $\sim$700 from 4.2 to 5.4$\mu$m. PAH and Aliphatic features are labeled. Calculated emission fits are in red.}.\label{fig3}
\end{figure}

\end{document}